\documentclass[letterpaper]{jpconf}
\usepackage{graphicx}
\begin{document}
\title{Faraday rotation in the MOJAVE blazars: 3C 273 a case study}

\author{T Hovatta$^{1,}$\footnote[2]{Present address:
Cahill Laboratory of Astronomy and Astrophysics, California Institute of Technology, 1200 E California Blvd, Pasadena CA 91125} , M L Lister$^1$, 
M F Aller$^3$, H D Aller$^3$, D C Homan$^4$, Y Y Kovalev$^{5,6}$, A B
Pushkarev$^{7,8}$ and T Savolainen $^6$}

\address{$^1$ Department of Physics, Purdue University, 525
  Northwestern Ave, West Lafayette, IN 47907, USA}
\address{$^3$ Department of Astronomy, University of Michigan, 817 Denison Building, Ann Arbor, MI 48109-1042, USA}
\address{$^4$ Department of Physics and Astronomy, Denison University, Granville, OH 43023, USA}
\address{$^5$ Astro Space Center of Lebedev Physical Institute,
  Profsoyuznaya 84/32, 117997 Moscow, Russia}
\address{$^6$ Max-Planck-Institut f\"ur Radioastronomie, Auf dem
  H\"ugel 69, 53121 Bonn, Germany}
\address{$^7$ Pulkovo Observatory, Pulkovskoe Chaussee 65/1, 196140
  St. Petersburg, Russia}
\address{$^8$ Crimean Astrophysical Observatory, 98409 Nauchny, Crimea, Ukraine}

\ead{thovatta@caltech.edu}

\begin{abstract}
Radio polarimetric observations of Active Galactic Nuclei can reveal
the magnetic field structure in the parsec-scale jets of these
sources. We have observed the $\gamma$-ray blazar 3C~273 as part of our multi-frequency survey with the Very Long Baseline
Array to study Faraday rotation in a large sample of jets. Our
observations re-confirm the transverse rotation measure gradient 
in 3C~273. For the first time the gradient is seen to cross zero which
is further indication for a helical magnetic field and spine-sheath
structure in the jet. We believe the difference to previous epochs is
due to a different part of the jet being illuminated in our observations.
\end{abstract}

\section{Introduction}
According to current knowledge, jets of AGN are magnetically launched from a
rotating black hole or accretion disk \cite{blandford77, meier01,
  vlahakis04}. It is then natural to expect the magnetic field
structure in the jet to be helical \cite{mckinney07}. We do not know,
however, if the helical magnetic field structure dominates down to
parsec-scale jets or if the field becomes predominately tangled due to re-collimation
shocks or interaction with the external medium
\cite{marscher08}. Magnetic field structure of parsec-scale AGN jets can be studied
by using polarimetric multifrequency observations. Evidence for
helical magnetic field structure could be a detection of a Faraday 
rotation measure (RM) gradient transverse to the jet
\cite{blandford93}. 

Faraday rotation occurs when polarized waves propagate through
non-relativistic plasma within or external to the source \cite{burn66}. 
It can be described with a linear dependence between the observed 
electric vector position angle (EVPA, $\chi_\mathrm{obs}$) and wavelength squared
($\lambda^2$) by the following formula 
\begin{equation}
\chi_\mathrm{obs} = \chi_0 + \frac{e^3\lambda^2}{8\pi^2\epsilon_0m^2c^3}\int n_e \mathbf{B} \cdot \mathbf{\mathrm{d}l} = \chi_0 + \mathrm{RM}\lambda^2,
\end{equation}
where $\chi_0$ is the intrinsic EVPA and RM is the rotation measure, related to the electron density $n_e$ and 
the magnetic field component $\mathbf{B}$ parallel to the line of
sight. The sign of the RM then depends on the direction of the
line of sight component of the magnetic field with RM being positive
if the magnetic field points towards the observer and negative if it
points away from the observer. Thus a helical magnetic field viewed
directly from the side would have a RM gradient going from positive to 
negative and crossing zero in the middle of the jet. Due to relativistic aberration effects 
blazars that are observed at small viewing angles are viewed from the side in the 
rest frame of the jet. Recent
simulations \cite{broderick10} show that launching the jet from a
rotating black hole can indeed produce such structure. 

The first observational evidence for a RM gradient across a jet was reported
in quasar 3C~273 by Asada et al. \cite{asada02}, who 
detected a RM change from 450 to 250 rad/m$^2$ across the jet in observations between 4.6 and 8.6 GHz in 1995. The
gradient was confirmed by Zavala \& Taylor 
\cite{zavala05} who in their two observing epochs in 2000 detected a
change from 2000 to 0 rad/m$^2$ across the 
jet of 3C~273 about 5 mas from the core. They attribute the large difference in the RM values in comparison to 
the previous observations \cite{asada02} to the better resolution of their high frequency observations between 12.1 and 22.2 GHz.
Follow-up observations on the gradient at frequencies between 4.6 and 8.6 GHz in 2002 reveal time variations on time
scales of several years \cite{asada08a}. Since these observations were done at the same angular
resolution as the first detection \cite{asada02}, the difference between the epochs cannot be 
due to observational effects. The total rotation they detect is larger than can be caused by internal Faraday rotation, as 
it would cause severe depolarization. They also rule out external clouds as the cause of the rotation as the variability 
time scales are too short to cause variations in foreground clouds. They conclude that the rotation is most likely caused 
by a sheath around the jet. RM gradients have been detected in other sources as well but the issue remains 
controversial due to the difficulty in assessing the reliability of a RM gradient \cite{taylor10}. Recently, 
RM gradients have been detected in the kpc-scale jets as well, see e.g. Gabuzda et al. (these 
proceedings).

In 2006 we conducted a large survey of Faraday rotation in 191 sources in
the MOJAVE (Monitoring of Jets in Active galactic nuclei with VLBA Experiments) 
sample (Hovatta et al. 2011 in preparation). In this paper we report
our observations of the RM gradient in 3C~273.

\section{Observations and data reduction}

We observed 3C~273 on March 9, 2006 and June 15, 2006 with the Very
Long Baseline Array (VLBA) at 8.1, 8.4, 12.1 and 15.3 GHz. The initial 
data reduction and calibration were done in the standard manner as 
described in the AIPS cookbook\footnote{http://www.aips.nrao.edu}. The
imaging and self calibration were done using the Difmap
package \cite{shepherd97}. For more details see Ref.~\cite{lister09}.
The absolute EVPA calibration of the 15 GHz observations
were done previously as part of the MOJAVE project using the stability of 
the D-term phases of some antennas and IFs \cite{gomez02,lister05}. The 8 and 12 GHz bands
were calibrated using nearby observations from the VLA/VLBA polarization calibration 
database\footnote{http://www.aoc.nrao.edu/$\sim$smyers/calibration/},
and single-dish observations from University of Michigan Radio
Astronomy Observatory. We estimate the uncertainty in the EVPA
calibration to be 3$^\circ$, 2$^\circ$ and 4$^\circ$ in the 15, 12 and
8\,GHz bands, respectively. 

The (u,v)-range of all the bands was clipped to correspond to each other,
and the images at all the frequencies were restored to the 8.1\,GHz beam. Because the absolute
position of the source is lost during our calibration process, we used
2D cross-correlation to properly align the images at different frequencies.
The images of the different bands were then combined to create 
the RM maps by performing a linear least-squares-fit of the EVPA 
vs. $\lambda^2$ in each pixel independently. 
Goodness of the fit was estimated by
a $\chi^2$ criterion. Details of the procedure will be reported in Hovatta
et al. 2011 (in preparation).

\section{RM maps of 3C~273}
\begin{figure}[h]
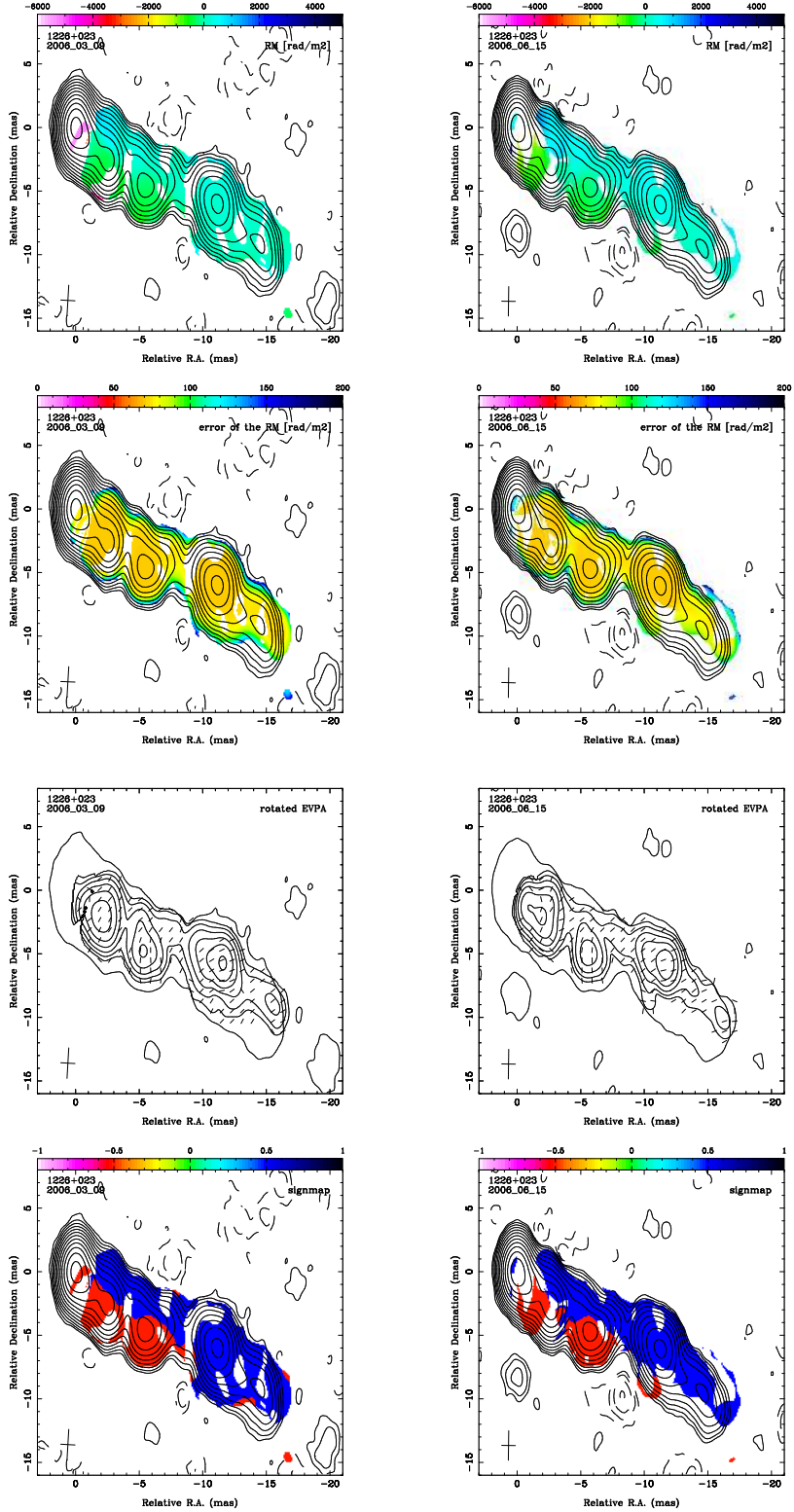

\begin{minipage}{14pc}
\includegraphics[width=14pc,angle=-90,scale=3.5]{1226+023.2006_03_09_rm_4panel.ps}
\end{minipage}
\begin{minipage}{14pc}
\includegraphics[width=14pc,angle=-90,scale=3.5]{1226+023.2006_06_15_rm_4panel.ps}
\end{minipage}
\begin{minipage}[b]{8pc}\caption{\label{rm_BL137B} Left: Epoch
    2006-03-09. Right: Epoch 2006-06-15. From top down: RM in color
    overlaid on 15 GHz total intensity contours, error of the RM, RM-corrected EVPA at 15 GHz overlaid on the 15 GHz polarized flux contours, sign of the RM. In the top panel the map has been blanked for very high absolute RM values.}
\end{minipage}
\end{figure}

The RM maps for 3C~273 are shown in Fig. \ref{rm_BL137B}. 
The top panels show the RM map with color indicating
the RM in rad$/\mathrm{m}^2$ overlaid on top of total intensity contours at 15
GHz. The second panel from the top shows the error in RM, calculated
from the variance-covariance matrix of the linear fit. The third panel
from the top shows the EVPAs at 15\,GHz which have been corrected for
the Faraday rotation. In the bottom panel the sign of the RM is shown.

The maps are blanked for pixels with absolute RMs of over 20 000 rad/m$^2$.
These are typically concentrated on the core or edge regions of the jets and
they do not coincide between the two epochs. Therefore we do not think
these extreme RM values are real but are artifacts that occur by chance due to 
random noise in the data and n$\pi$ ambiguities present in the analysis. 
We are currently investigating this in detail with simulations 
(Hovatta et al. 2011, in preparation),
and our preliminary results show that it is possible to get these large RM values purely due to 
random noise. The errors in the RM are typically around 70
rad/m$^2$ rising to as high as 200 rad/m$^2$ in the jet edges where the signal-to-noise 
ratio in total intensity is typically smaller. 
The Faraday-corrected EVPAs are mostly perpendicular to the optically
thin jet indicating that the magnetic field is predominantly parallel
to the jet. The jet region about 7 mas from the core is an exception
with clearly differing EVPA direction. This could be caused by a shock
compressing the magnetic field which would also explain the enhanced
total intensity and polarized emission from this region. 
The sign map
of the RM shows clearly bilateral structure indicative of a helical
magnetic field as shown by simulations \cite{broderick10}.

\section{Transverse RM gradient}
\begin{figure}
\begin{center}
\includegraphics[scale=0.37]{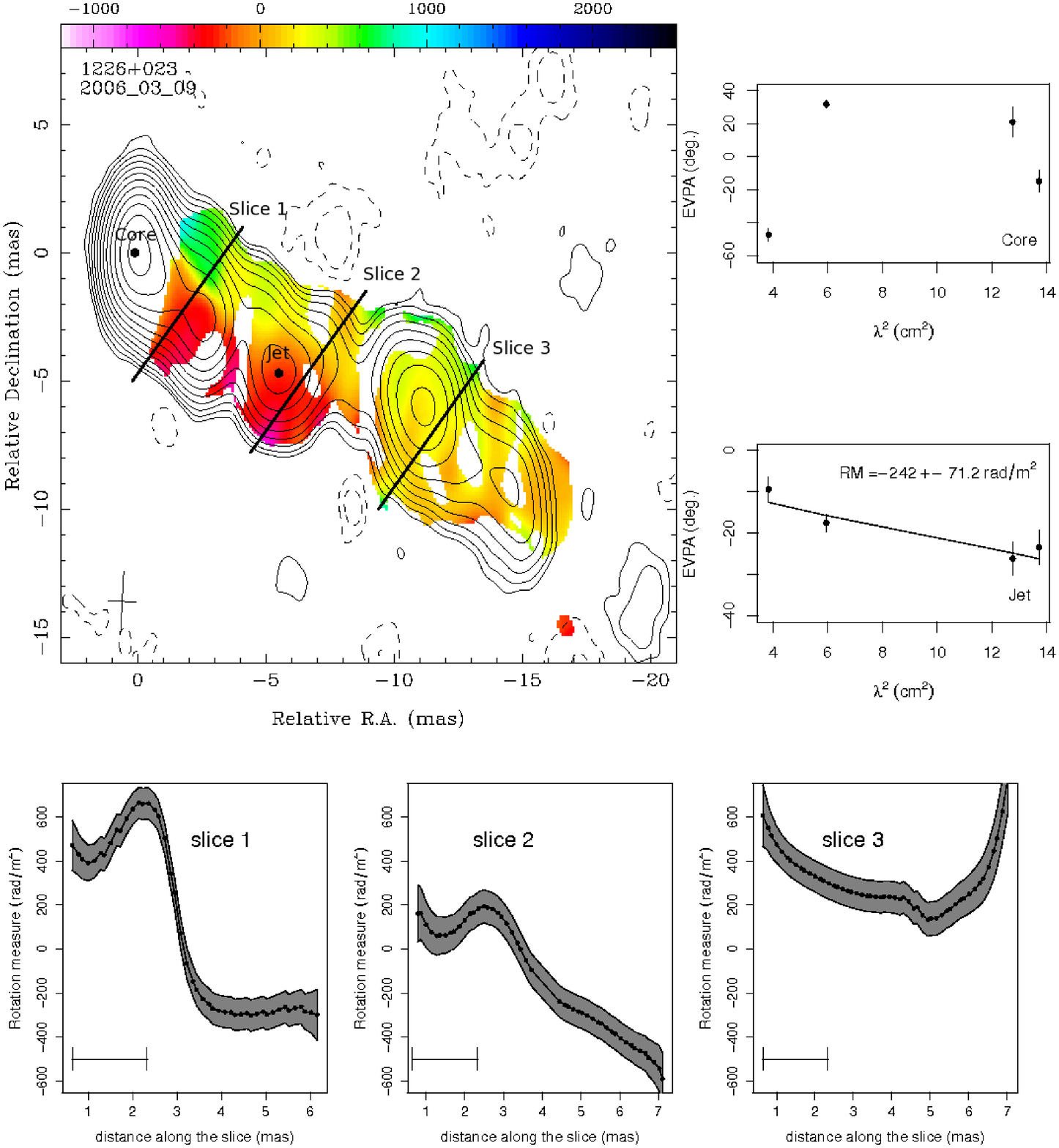}
\end{center}
\caption{\label{grad_BL137B}Left: RM map of 3C~273 in 2006-03-09 with
  additional high-RM pixels of over 4000 rad/m$^2$ blanked to show the gradient. Right: RM fit of the
  core and a jet component 7 mas from the core as marked by the dots in the RM map. Bottom: Transverse
  slices of the jet as marked by the lines in the RM map. The absolute EVPA calibration error has been subtracted from the error bars because it is the same in every pixel.
The beam size along the slice is shown as a line in the bottom left corner.}
\end{figure}

\begin{figure}
\begin{center}
\includegraphics[scale=0.37]{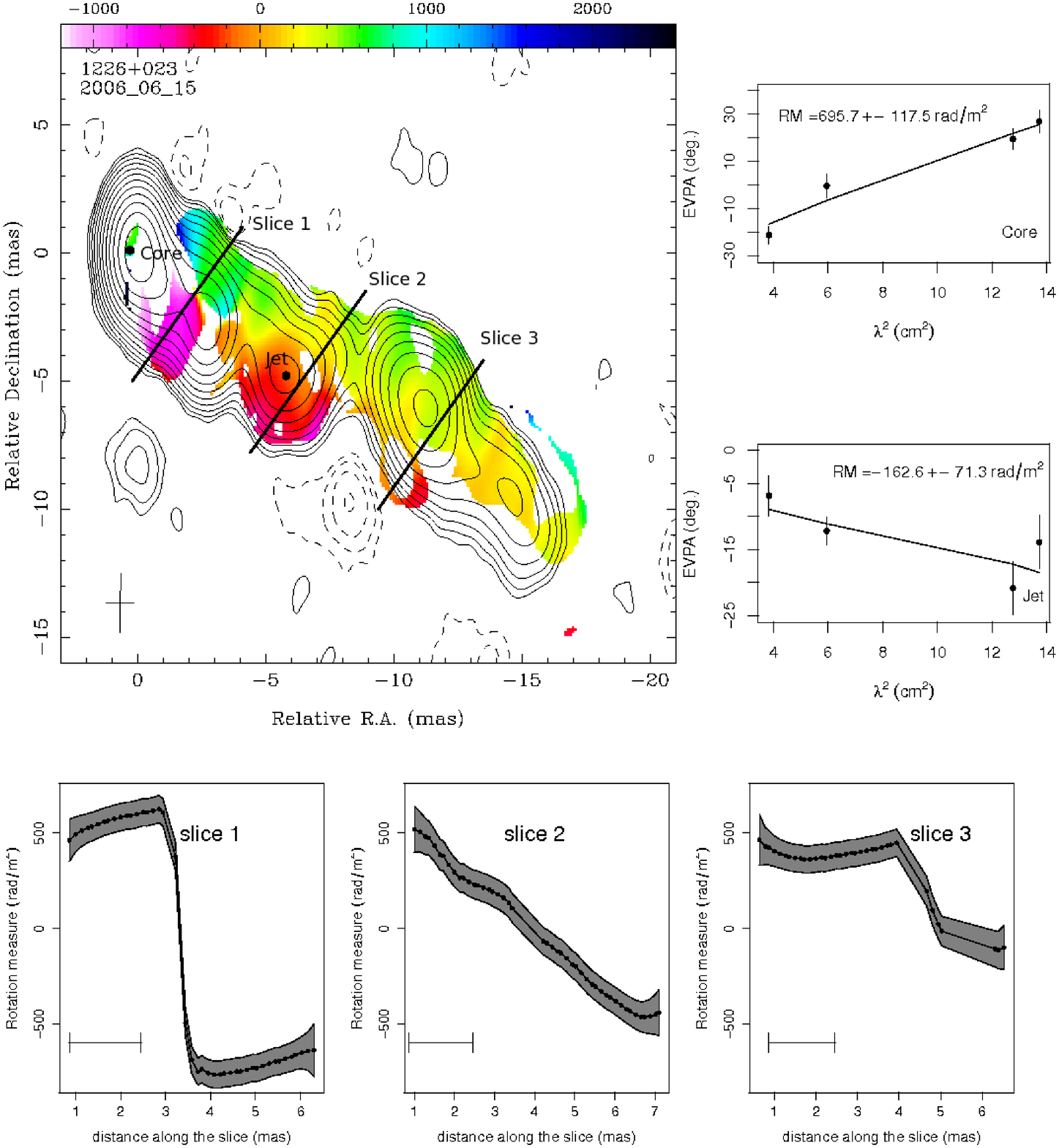}
\end{center}
\caption{\label{grad_BL137F}Same as Fig. \ref{grad_BL137B} but for epoch 2006-06-15.}
\end{figure}

Our observations in Figs. \ref{grad_BL137B} and
\ref{grad_BL137F} show a clear gradient transverse to the jet and along the whole jet length. The maximum gradient is seen about 3-7 mas 
from the core where the RM changes from 500 to -600 rad/m$^2$. Moving down the jet the gradient becomes 
less pronounced as was also noted by Attridge et al.\cite{attridge05}. Our observations were done at resolution which is in
between those by Refs.~\cite{asada02, asada08a} and Ref.~\cite{zavala05} and that can explain some of the discrepancy in our values. The most notable difference is that our 
RM gradient goes from positive to negative RM values, as expected for
a helical field viewed from the side, in contrast to Refs.~\cite{asada02,asada08a, zavala05} where the values are always positive. Our positive 
values also do not extend to as high values as in the maps of Ref.~\cite{zavala05}. The most likely explanation is 
that we are seeing a different part of the jet being illuminated by components moving down the jet as our observations 
are several years apart from earlier observations \cite{savolainen06}. The effect of seeing a different part of the jet at different epochs is 
clearly demonstrated in the stacked images of the MOJAVE program \cite{lister09} which show how the jet at any given epoch 
may fill only a portion of the region visible in the stacked image. In fact, if we calculate the 
inner jet direction in 2000 and compare it to the jet direction in 2006, the difference is 10$^\circ$. We are thus seeing a 
portion of the jet which is 10$^\circ$ to the South of the observations in Ref.~\cite{zavala05}. This together with the slightly 
different resolution of our observations could explain the difference in the RM values. A similar effect is 
seen in the RM structure of the radio galaxy 3C~120 over several years time scales \cite{gomez08, gomez11}.

Even though our observations are only three months apart we can still see differences in the RM between the two 
epochs. In particular, the core regions strongly differ. This is probably due 
to the complex sub-structure of the core region in 3C~273
\cite{attridge05, savolainen08}. If we ignore the unreliable jet edges, the jet RM values are very similar 
in the two epochs and differences are likely due to observational effects such as signal to noise in 
the differing jet regions resulting in bad $\lambda^2$-fits in one epoch or the other to be in different 
parts of the jet. The gradient looks very similar over the two
epochs. 
The consistent RM gradient over the length of the whole jet suggests a scenario involving
 a sheath around the jet. 
 Recent simulations show that observing a bilateral structure in the sign map
 of the RM is a strong indication of helical magnetic field
 surrounding the jet which cannot be easily explained by a random
 foreground screen \cite{broderick10}.

Detailed multi-epoch observations of the RM and polarization structure are needed 
to fully understand the Faraday-screen properties, as was recently done for
3C~120 \cite{gomez11}. Their observations 
showed that even though they detect a gradient and variations in the jet RM, these are not connected with the 
polarization changes. This suggests that the Faraday medium may be a distant external screen not connected with 
the jet.

\section{Conclusions}
We have observed 3C~273 with the VLBA as part of our large survey to
study the Faraday rotation in AGN jets. Our observations confirm
the rotation measure gradient transverse to the jet. For the first
time the gradient is seen to cross zero which is further evidence for
a helical magnetic field and a sheath around the jet. 
\scriptsize{
\ack
The MOJAVE project is supported under National Science Foundation grant AST-
0807860 and NASA Fermi grant NNX08AV67G. Work at UMRAO has been supported 
by a series of grants from the NSF and NASA and by funds
for operation from the University of Michigan.The VLBA is a facility of the 
National Science Foundation operated by the National
Radio Astronomy Observatory under cooperative agreement with Associated Universities,
Inc.}


\end{document}